\documentclass[12pt]{article}
\usepackage{epsfig}
\usepackage{hyperref}%

\date{}
\begin{document}

\newcommand{\beq}{\begin{equation}}
\newcommand{\eeq}{\end{equation}}
\newcommand{\nn}{\nonumber}
\newcommand{\bea}{\begin{eqnarray}}
\newcommand{\eea}{\end{eqnarray}}

\title{Braneworlds, Conformal Fields and Dark Energy}

\author{Rui Neves\\
{\small \it Departamento de F\'{\i}sica, Faculdade de
 Ci\^encias e Tecnologia}\\ 
{\small \it Universidade do Algarve}\\{\small \it \&}\\{\small \it 
Centro Multidisciplinar de
 Astrof\'{\i}sica-CENTRA}\\
{\small \it Campus de Gambelas, 8005-139 Faro, Portugal}\\
{\small \it  E-mail: rneves@ualg.pt}
}

\maketitle

\begin{abstract}

In the Randall-Sundrum scenario we analize the dynamics of
a spherically symmetric 3-brane when matter fields propagate in the bulk. For a well defined class of
conformal fields of weight -4 we determine a new set of exact 5-dimensional solutions which localize gravity in the
vicinity of the brane and are stable under radion field
  perturbations. Geometries which describe the dynamics of inhomogeneous dust, generalized dark radiation and 
homogeneous polytropic dark energy on the brane are shown to belong to this set.
\end{abstract}

\section{Introduction}

In the Randall-Sundrum (RS) scenario \cite{RS1,RS2} the
visible Universe is a 3-brane world of a $Z_2$ symmetric 
5-dimensional (5D) anti-de Sitter (AdS) space. In the RS1 model
\cite{RS1} there is a compact fifth dimension and two brane
boundaries. The gravitational field is localized near the hidden positive
tension brane and decays towards the visible negative tension brane. 
In this model the 
hierarchy problem is reformulated as an exponential hierarchy between the weak
and Planck scales \cite{RS1}. In the RS2 model \cite{RS2} there is a
single positive tension brane in an infinite fifth dimension. Then
gravity is bound to the positive tension brane now interpreted as the
visible brane. 

At low energies the theory of gravity on the observable brane is 
4-dimensional (4D) general relativity and the cosmology may be 
Friedmann-Robertson-Walker \cite{RS1}-\cite{TM}. 
In the RS1 model this is only possible if  
the radion mode is stabilized using for example a bulk scalar
field \cite{GW,WFGK,CGRT,TM}. Gravitational 
collapse was also analyzed in the RS scenario \cite{CHR}-\cite{RC2}. 
However, an exact 5D solution representing a stable black hole
localized on a 3-brane has not yet been discovered.   
So far, the only known static black holes localized
on a brane remain to be those found for a 2-brane in a 4D AdS space
\cite{EHM}. A solution to this problem requires the non-singular
localization of both gravity and matter \cite{CHR,KOP}-\cite{RC2} and could be connected to quantum black holes on the brane \cite{EFK}. This is an extra motivation to
look for 5D collapse solutions localized on a
brane. In addition, the effective covariant
Gauss-Codazzi approach \cite{BC,SMS} has permitted the discovery of
many braneworld solutions which have not yet been associated with
exact 5D spacetimes \cite{COSp}-\cite{RC1}.

In this paper we continue the research
on the dynamics of a spherically symmetric RS 3-brane when 5D
conformal matter fields propagate in the bulk \cite{RC2,RC3} (see also
\cite{EONO}). In our previous work
\cite{RC2,RC3} we have found a new class of exact
5D dynamical solutions for which gravity is localized near
the brane by the exponential RS warp. These solutions were shown to describe on the brane the dynamics of
inhomogeneous dust, generalized dark radiation and homogeneous
polytropic matter. However, the density and pressures
of the conformal bulk fluid increase with the coordinate of the fifth
dimension. In the RS2 model this generates a divergence at the AdS
horizon as in the Schwarzschild black string solution \cite{CHR}. In the RS1 model this is not a problem. However, the solutions turn out to be unstable under radion field 
perturbations \cite{RN}. In this
work we report on a new set of exact 5D braneworld solutions which have a stable radion mode and
still describe on the observable brane the dynamics of
inhomogeneous dust, generalized dark radiation and homogeneous polytropic matter. 

\section{5D Einstein Equations and Conformal Fields}

On the 5D RS orbifold the non-factorizable metric consistent with the $Z_2$ symmetry in $z$ and
with 4D spherical symmetry on the brane corresponds to the 
5D line element
$d{\tilde{s}_5^2}={\Omega^2}\left(d{z^2}-{e^{2A}}d{t^2}+{e^{2B}}d{r^2}+{R^2}d{\Omega_2^2}\right)$,
where $\Omega=\Omega(t,r,z)$, $A=A(t,r,z)$, $B=B(t,r,z)$ and
$R=R(t,r,z)$ are $Z_2$ symmetric functions. $R(t,r,z)$ represents the
physical radius of the 2-spheres and $\Omega$ is the warp factor which
defines a global conformal transformation on the metric.

The classical dynamics is defined by the 5D Einstein equations,
\beq
{\tilde{G}_\mu^\nu}=-{\kappa_5^2}\left\{{\Lambda_B}{\delta_\mu^\nu}+
{1\over{\sqrt{\tilde{g}_{55}}}}\left[\lambda\delta
\left(z-{z_0}\right)+\lambda'\delta\left(z-{{z'}_0}\right)\right]\left({\delta_\mu^\nu}-{\delta_5^\nu}
{\delta_\mu^5}\right)-{\tilde{\mathcal{T}}_\mu^\nu}\right\},
\label{5DEeq}
\eeq
where $\Lambda_B$ is the negative bulk cosmological constant,
$\lambda$ and $\lambda'$ are the brane tensions, ${\kappa_5^2}=8\pi/{M_5^3}$ with $M_5$ the
fundamental 5D Planck mass and ${\tilde{\mathcal{T}}_\mu^\nu}$ is
the stress-energy tensor associated with the matter fields. In 5D ${\tilde{\mathcal{T}}_\mu^\nu}$ is conserved,
${\tilde{\nabla}_\mu}{\tilde{\mathcal{T}}_\nu^\mu}=0$.

Let us consider the special class of conformal bulk matter defined by ${\tilde{\mathcal{T}}_\mu^\nu}={\Omega^{-2}}{\mathcal{T}_\mu^\nu}$ and
assume that $\mathcal{T}_\mu^\nu$ depends only on $t$ and $r$. The conformal stress-energy tensor ${\tilde{\mathcal{T}}^{\mu\nu}}$
may be separated in two sectors
${\tilde{T}^{\mu\nu}}$ and ${\tilde{U}^{\mu\nu}}$ with the same weight $s$,
${\tilde{\mathcal{T}}^{\mu\nu}}={\tilde{T}^{\mu\nu}}+{\tilde{U}^{\mu\nu}}$
where ${\tilde{T}^{\mu\nu}}={\Omega^s}{T^{\mu\nu}}$ and
${\tilde{U}^{\mu\nu}}={\Omega^s}{U^{\mu\nu}}$. Assuming that $T_\mu^\nu$ and $U_\mu^\nu$ are conserved tensor fields, $A=A(t,r)$, $B=B(t,r)$, $R=R(t,r)$ and $\Omega=\Omega(z)$ we obtain
\beq
{G_a^b}={\kappa_5^2}{T_a^b},\quad{G_5^5}={\kappa_5^2}{T_5^5}\quad{\nabla_a}{T_b^a}=0,\quad {\nabla_a}{U_b^a}=0,\label{4DECeq}
\eeq
\beq
6{\Omega^{-2}}{{({\partial_z}\Omega)}^2}+{\kappa_5^2}{\Omega^2}{\Lambda_B}={k_5^2}{U_5^5}\label{rswf1}
\eeq
\beq
\left\{3{\Omega^{-1}}{\partial_z^2}\Omega+{\kappa_5^2}{\Omega^2}
\left\{{\Lambda_B}+{\Omega^{-1}}\left[\lambda\delta(z-{z_0})+\lambda'\delta(z-{{z'}_0})\right]\right\}\right\}{\delta_a^b}={k_5^2}{U_a^b},\label{rswf2}
\eeq
where the latin indices represents the 4D coordinates
$t$, $r$, $\theta$ and $\phi$. On the other hand we also  
find the following equations of state 
$2{T_5^5}={T_c^c}, 2{U_5^5}={U_c^c}$.
Note that $U_\mu^\nu$ must be a diagonal tensor field,
${U_\mu^\nu}=\mbox{diag}\left(-\bar{\rho},{\bar{p}_r},{\bar{p}_T},{\bar{p}_T},{\bar{p}_5}\right)$
with constant density and pressures satisfying 
$\bar{\rho}=-{\bar{p}_r}=-{\bar{p}_T}$ and ${\bar{p}_5}=-2\bar{\rho}$. On the other hand if ${T_\mu^\nu}=\mbox{diag}\left(-\rho,{p_r},{p_T},{p_T},{p_5}\right)$ where $\rho$, $p_r$, $p_T$ and $p_5$ denote bulk matter density and
pressures then its equation of state is re-written as
\beq
\rho-{p_r}-2{p_T}+2{p_5}=0,\label{eqst3}
\eeq
where $\rho$, $p_r$, $p_T$ and $p_5$ must be independent of
$z$ but may be functions of $t$ and $r$. The bulk matter is,
however, inhomogeneously distributed along the fifth dimension because
the physical energy density, ${\tilde \rho}(t,r,z)$, and pressures,
$\tilde{p}(t,r,z)$, are related
to $\rho(t,r)$ and $p(t,r)$ by the scale factor $\Omega^{-2}(z)$. Note
also that the warp depends on the
conformal bulk fields only through $U_\mu^\nu$. So the role of $U_\mu^\nu$ is to influence how
the gravitational field is warped around the
branes. On the other hand $T_\mu^\nu$ determines the dynamics on the
branes. In the RS1 model the two branes have identical cosmological
evolutions and gravity will be localized on the Planck brane and not on the
visible one. 

\section{Exact 5D Warped Solutions}

The dynamics of the $AdS_5$ braneworlds is defined by the solutions of equations (\ref{4DECeq}) to
(\ref{eqst3}). Let us first solve
the warp equations (\ref{rswf1}) and
(\ref{rswf2}). If ${\bar{p}_5}=0$ then $U_\mu^\nu=0$ and we obtain the usual RS warp equations. A solution is the exponential RS warp \cite{RS1,RS2}. Using the coordinate $y$ related to $z$ by
$z=l{e^{y/l}}$ for $y>0$ we find  
\beq
\Omega(y)={\Omega_{\mbox{\tiny
      RS}}}(y)={e^{-|y|/l}},\label{rswf}
\eeq
where $l$ is the AdS radius given by
$l=1/\sqrt{-{\Lambda_B}{\kappa_5^2}/6}$.
If $\bar{p}_5$ is non-zero then there is a new set of warp solutions
to be considered. Integrating Eq. (\ref{rswf1}) and taking into
account the $Z_2$ symmetry we find 
\beq
\Omega(y)={e^{-|y|/l}}\left(1+{{\bar{p}_5}\over{4{\Lambda_B}}}{e^{2|y|/l}}\right).\label{wfp5y}
\eeq
This set of solutions which depends on the 5D pressure
$\bar{p}_5$ must also satisfy Eq. (\ref{rswf2}) which contains the 
Israel conditions. This may only happen if the brane tensions
$\lambda$ and $\lambda'$ are given by
\beq
\lambda={6\over{l{\kappa_5^2}}}{{1-{{\bar{p}_5}\over{4{\Lambda_B}}}}\over{1+{{\bar{p}_5}\over{4{\Lambda_B}}}}},\quad \lambda'=-{6\over{l{\kappa_5^2}}}{{1-{{\bar{p}_5}\over{4{\Lambda_B}}}{e^{2\pi{r_c}/l}}}\over{1+{{\bar{p}_5}\over{4{\Lambda_B}}}{e^{2\pi{r_c}/l}}}},\label{wft4}
\eeq
where $r_c$ is the RS compactification scale.

The conformal factor $\Omega(y)$ defines how the gravitational field is warped
around the brane. To find the dynamics on the brane
we need to consider solutions of Eq. (\ref{4DECeq}) when the diagonal bulk matter $T_\mu^\nu$ 
satisfies Eq. (\ref{eqst3}). 
For inhomogeneous dust, generalized dark
radiation and homogeneous polytropic matter such solutions were
determined in Refs. \cite{RC2} and \cite{RC3}. The latter describes the
dynamics on the brane of dark energy in the form of a polytropic
fluid. The diagonal conformal matter may be defined by
$\rho={\rho_{\mbox{\tiny P}}}, {p_r}+\eta{{\rho_{\mbox{\tiny
P}}}^\alpha}=0, {p_T}={p_r}, {p_5}=-\left({\rho_{\mbox{\tiny P}}}
+3\eta{{\rho_{\mbox{\tiny P}}}^\alpha}\right)/2$, where ${\rho_{\mbox{\tiny P}}}$ defines the polytropic energy
density and the parameters ($\alpha$, $\eta$) characterize
different polytropic phases. For $-1\leq\alpha<0$ the fluid is in its
generalized Chaplygin phase (see also \cite{BBS}). The 5D polytropic solutions are \cite{RC3}
\beq
d{\tilde{s}_5^2}={\Omega^2}\left[-d{t^2}+{S^2}
\left({{d{r^2}}\over{1-k{r^2}}}+{r^2}d{\Omega_2^2}\right)\right]+d{y^2},
\label{dmsol1}
\eeq
where the Robertson-Walker brane scale factor $S$ satisfies 
\beq
{\dot{S}^2}=-k+{{\kappa_5^2}\over{3}}{S^2}{{\left(\eta+{a\over{S^{3-3\alpha}}}\right)}^
{1\over{1-\alpha}}}.
\eeq

\section{Radion Stability}

To analyze how these solutions behave under radion field
perturbations we apply a saddle point expansion procedure
based on the action \cite{HKP,CGHW}. Let us write the most general metric
consistent with the 
$Z_2$ symmetry in $y$ and with 4D spherical symmetry on the
brane in the form
$d{\tilde{s}^2}={a^2}d{s_4^2}+{b^2}d{y^2}$ with $d{s_4^2}=-d{t^2}
+{e^{2B}}d{r^2}+{R^2}d{\Omega_2^2}$. The metric functions $a=a(t,r,y)$, $B=B(t,r,y)$,
$R=R(t,r,y)$ and $b=b(t,r,y)$ are $Z_2$ symmetric. Now $a$ is the warp
factor and $b$ is related to the radion field. 
The 5D dynamical RS action is given by
\beq
\tilde{S}=\int{d^4}xdy\sqrt{-\tilde{g}}
\left\{{\tilde{R}\over{2{\kappa_5^2}}}-
{\Lambda_B}-{1\over{\sqrt{\tilde{g}_{55}}}}\left[\lambda\delta\left(y\right)+\lambda'\delta\left(y-\pi
  {r_c}\right)\right]+
{\tilde{L}_B}\right\}.\label{5Dact1}
\eeq
Our braneworld backgrounds correspond to the metric functions $b=1$,
$B=B(t,r)$, $R=R(t,r)$ and $a=\Omega(y)$.

To calculate the radion potential we consider the dimensional
reduction of (\ref{5Dact1}). Using the metric with $a(t,r,y)=\Omega(y){e^{-\beta(t,r)}}$ and
$b(t,r)={e^{\beta(t,r)}}$ we obtain in the Einstein frame \cite{RN}
\beq
\tilde{S}=\int{d^4}x\sqrt{-{g_4}}\left({{R_4}\over{2{\kappa_4^2}}}-{1\over{2}}{\nabla_c}\gamma{\nabla_d}\gamma{g_4^{cd}}-\tilde{V}\right),\label{DR5Dact}
\eeq
where $\gamma=\beta/({\kappa_4}\sqrt{2/3})$ is the canonically
normalized radion field. The function $\tilde{V}=\tilde{V}(\gamma)$ is
the radion potential and it is given by
\[
\tilde{V}={2\over{\kappa_5^2}}{\chi^3}\left[3
\int dy{\Omega^2}
{{({\partial_y}\Omega)}^2}+2\int dy{\Omega^3}
{\partial_y^2}\Omega\right]+\chi\int dy{\Omega^4}\left({\Lambda_B}-{\tilde{L}_B}\right)
\]
\beq
+{\chi^2}\int dy{\Omega^4}\left[\lambda\delta\left(y\right)+\lambda'\delta\left(y-\pi
  {r_c}\right)\right],\label{rp}
\eeq
where the field $\chi$ is defined as
$\chi={e^{-\sqrt{(2/3)}\;{\kappa_4}\gamma}}$. The integration in the fifth dimension is performed in the interval
$\left[-\pi{r_c},\pi{r_c}\right]$ and that we have chosen 
$\int dy{\Omega^2}={\kappa_5^2}/{\kappa_4^2}$.

To analyze the stability of the $AdS_5$ braneworld solutions we consider a
saddle point expansion of the radion field potential $V$. If
${\bar{p}_5}=0$ then $\Omega={\Omega_{\mbox{\tiny RS}}}$. The critical
extremum corresponding to our braneworlds is $\chi=1$ \cite{RN}. 
Stable solutions must be associated with a positive second variation of the radion potential. 
If the equation of state of the conformal bulk fields is independent
of the radion perturbation then for $\chi=1$ 
the second variation is negative and so the corresponding braneworlds are unstable \cite{RN}. If the equation of state is kept invariant
under the radion perturbations it is possible to find stable solutions
at $\chi=1$ if the warp is changed. Indeed, the new
relevant warp functions are given in Eq. (\ref{wfp5y}) and stability
exists for a range of the parameters if ${\bar{p}_5}<0$. As an
example consider the interval 
$4{\Lambda_B}{e^{-2\pi{r_c}/l}}<{\bar{p}_5}<0$ which corresponds to a
brane configuration with $\lambda>0$ and $\lambda'<0$. Then the stability
conditions are
$l>3{r_c},4{\Lambda_B}{e^{-2\pi{r_c}/l}}<{\bar{p}_5}<{\bar{p}_5^*}(l)$ 
and
${\bar{p}_5^*}(l)<{\Lambda_B}/2$.

\section{Conclusions}

In this paper we have analized exact
5D dynamical solutions with gravity localized near
the brane which are associated with conformal bulk fields of weight -4
and describe the dynamics of inhomogeneous dust, generalized dark
radiation and homogeneous
polytropic matter on the brane. We have discussed their behaviour under
radion field perturbations and shown that they are extrema of the
radion potential. We have also shown
that if the equation of state characterizing the conformal fluid is
independent of the perturbation then the radion may be stabilized by a sector of the conformal fields while
    another sector generates the dynamics on the brane. Stabilization requires a bulk fluid with a constant
    negative pressure and involves new warp functions. On the brane these solutions also describe the dynamics of
inhomogeneous dust, generalized dark radiation and homogeneous
polytropic matter. Whether
    gravity is suficiently localized on the brane is an open
    problem for future research.

\section*{Acknowledgments}

We would like to thank the financial support of {\it Funda\c {c}\~ao para a Ci\^encia e a Tecnologia} (FCT) and {\it Fundo Social Europeu} (FSE) under the contract
SFRH/BPD/7182/2\\001 ({\it III Quadro Comunit\'ario de Apoio}) as well
as of {\it Centro
  Multidisciplinar de Astrof\'{\i}sica} (CENTRA).

\end{document}